\documentclass[%
 aip,apl,
 amsmath,amssymb,
 reprint,%
]{revtex4-1}

\usepackage{graphicx}
\usepackage{dcolumn}
\usepackage{bm}

\usepackage[utf8]{inputenc}
\usepackage[T1]{fontenc}
\usepackage{mathptmx}

\begin{document}


\title{Spin-transport in superconductors}

\author{K.~Ohnishi}%
\affiliation{ 
Department of Physics, Kyushu University, 744 Motooka, Fukuoka, 819-0395, Japan
}%
\affiliation{ 
Research center for Quantum Nano-Spin Sciences, 744 Motooka, Fukuoka, 819-0395, Japan
}%
\author{S.~Komori}%
\affiliation{ 
Department of Materials Science \& Metallurgy, University of Cambridge, 27 Charles Babbage
Road, Cambridge CB3 0FS, United Kingdom
}%
\author{G.~Yang}%
\affiliation{ 
Department of Materials Science \& Metallurgy, University of Cambridge, 27 Charles Babbage
Road, Cambridge CB3 0FS, United Kingdom
}%
\author{K.-R.~Jeon}%
\affiliation{ 
Department of Materials Science \& Metallurgy, University of Cambridge, 27 Charles Babbage
Road, Cambridge CB3 0FS, United Kingdom
}%
\affiliation{ 
Max Planck Institute of Microstructure Physics, Weinberg 2, 06120 Halle (Saale), Germany
}%
\author{L.~A.~B.~Olde Olthof}%
\affiliation{ 
Department of Materials Science \& Metallurgy, University of Cambridge, 27 Charles Babbage
Road, Cambridge CB3 0FS, United Kingdom
}%
\author{X.~Montiel}%
\affiliation{ 
Department of Materials Science \& Metallurgy, University of Cambridge, 27 Charles Babbage
Road, Cambridge CB3 0FS, United Kingdom
}%
\author{M.~G.~Blamire}%
\affiliation{ 
Department of Materials Science \& Metallurgy, University of Cambridge, 27 Charles Babbage
Road, Cambridge CB3 0FS, United Kingdom
}%
\author{J.~W.~A.~Robinson}%
 \email{jjr33@cam.ac.uk}
\affiliation{ 
Department of Materials Science \& Metallurgy, University of Cambridge, 27 Charles Babbage
Road, Cambridge CB3 0FS, United Kingdom
}%

\date{\today}

\begin{abstract}
Spin-transport in superconductors is a subject of fundamental and technical importance with
the potential for applications in superconducting-based cryogenic memory and logic.
Research in this area is rapidly intensifying with recent discoveries establishing the field of
superconducting spintronics. In this perspective we provide an overview of the experimental
state-of-the-art with a particular focus on local and nonlocal spin-transport in
superconductors, and propose device schemes to demonstrate the viability of
superconducting spin-based devices.
\end{abstract}

\maketitle

Spintronics has developed as an extension of semiconductor transistor based electronics by
utilising both the spin and charge degrees of freedom of electrons. Spintronic devices rely on
the creation and transfer of spin-angular momentum through spin current transport in
magnetic and nonmagnetic materials\cite{1,2}. Ferromagnets are key materials for spin
generation, detection and manipulation in semiconductors along with normal (nonmagnetic)
metals for transmitting spin information between ferromagnetic elements in a circuit.
However, equivalent experiments in which the normal metal is replaced with a
superconductor such as Al or Nb are far less understood. In a standard s-wave
superconductor, the electrons pair (``Cooper pairs'') with opposite spins in a so-called spinsinglet
state. Singlet pairs and thus singlet supercurrents do not carry a net spin and hence
spin transmission in the superconducting state depends on the superconducting gap and is
also highly sensitive to the magnetisation structure of the proximitized ferromagnet\cite{3,4,5}.

The conflict between superconductivity and ferromagnetism results in intriguing spin-related
phenomena at $s$-wave superconductor/ferromagnet (SC/FM) interfaces\cite{5,6,7,8,9}. Supercurrents
are dissipationless and Cooper pairs are coherent and hence, combing superconductivity and
spin-related phenomena offers the potential for new devices. These device may offer an
energy efficient means of generating spin-currents\cite{7,8,9} and also exciting avenues for spindependent
quantum phenomena\cite{8,10} as well as the discovery of quantum materials\cite{10},
and energy efficient computing\cite{7,8,11,12}. In this perspective, we survey recent advances
and outstanding issues in the area of spin-dependent transport in superconductors, and
outline device concepts that could exploit spin supercurrents in active devices.

\begin{figure}
\centering
\includegraphics[bb=0 0 1396 404, scale=0.23]{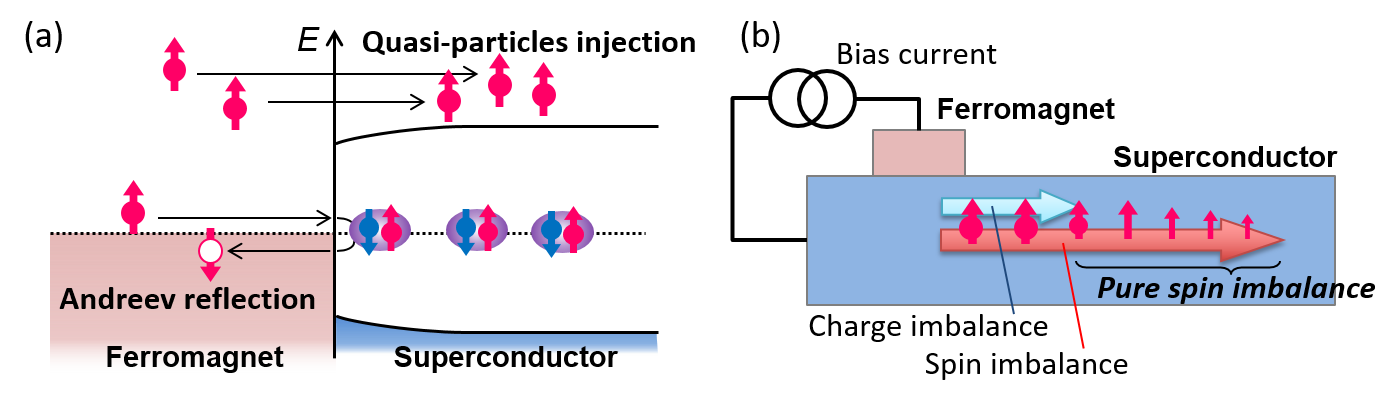}
\caption{(a) Spin-polarised electrons at a ferromagnet (FM) / superconductor (SC) interface. The
injected electrons induce a spin-polarisation of quasiparticles while Andreev reflection lowers
the transmission of the spin component of the current into the superconductor. (b) Injected
spin-current is propagated in a lateral structure as a pure spin current.}
\label{fig1}
\end{figure}

Quasiparticle (i.e. unpaired electrons) spin currents have been investigated through spin
current injection experiments into superconductors, both in local\cite{12,13,14,15,16,17,18,19,20,21,22} 
and non-local\cite{23,24,25,26} device geometries. At a FM/SC interface, an electron injected from FM can either enter
the SC as a quasiparticle with an energy above the superconducting gap or form a Cooper pair
via Andreev reflection\cite{3,27}. As shown in Fig.~\ref{fig1}(a), since singlet pairs cannot carry a net spinangular
momentum, spin-polarised electrons are injected as quasi-particles in a SC with a
voltage bias\cite{13,14,15,16,18,19,20,21,22} and/or thermal excitation\cite{23,24,25}.

An injected current from a FM induces a non-equilibrium accumulation of charge and spin at
a SC/FM interface, which results in a charge and spin imbalance within the SC near the SC/FM
interface\cite{5,28}. Since the transport equations for spin and charge imbalances are different
in a SC, the decay length of spin imbalance can be longer than that of charge imbalance\cite{21,29}.
Therefore, the diffusion of spin-imbalance (i.e. spin diffusive currents) without a charge
imbalance is possibly away from the junction point as illustrated in Fig.~\ref{fig1}(b). This can be
termed as a ``pure spin-imbalance'' as an analogue of pure spin currents which are a spin
current without a net charge current in NM. The pure spin-imbalance was detected as a 	spin
signal using a nonlocal FM probe -- e.g. F.~H$\rm\ddot{u}$bler et.~al.\cite{20} and C.~H.~L.~Quay et.~al.\cite{21} reported
evidence for spin-polarised quasiparticle diffusive currents in a lateral spin-valve using
superconducting Al as the spin-current channel and injected the spin-polarised current from
a FM into the Zeeman-spin-split superconducting density of states in Al.

The spin diffusive current in a SC is described by diffusion equations with key parameters
including the spin-polarisation and spin-diffusion time/length\cite{2,9,30}. To manipulate a spin
diffusive current in a SC and utilise the spin for applications, these parameters should be
carefully investigated and understood. They have been estimated through techniques
including the Hanle\cite{19,31} and spin Hall\cite{25,32} effects, and the local/nonlocal spin signals in
spin-valve structure\cite{19,24,26,30}. Despite intensive investigation of the spin diffusion current
within SCs, the relationship between spin-relaxation times and diffusion lengths remains
unclear\cite{32,33} and much debated.

\begin{figure}
\centering
\includegraphics[bb=0 0 1326 418, scale=0.23]{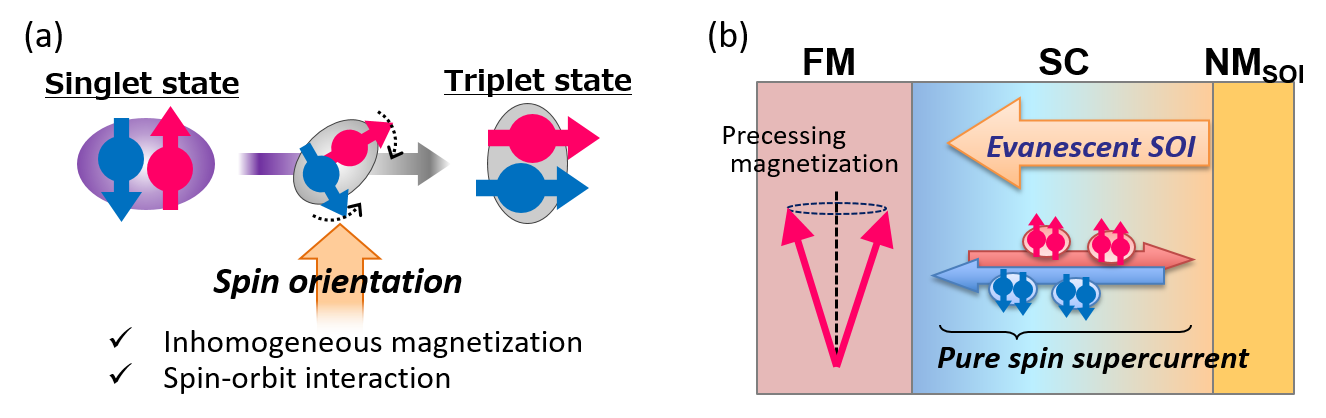}
\caption{(a) Schematic illustration showing conversion of singlet pairs to spin-polarized triplet
pairs. (b) Spin-polarized supercurrent in a superconductor (SC) in contact with a normal
metal (NM) with strong spin-orbit interaction (SOI) and a ferromagnet (FM).}
\label{fig2}
\end{figure}

A second potential type of spin current is the condensate spin current (so-called spin
supercurrent), which is carried by spin-triplet Cooper pairs. Spin triplet Cooper pairs are spinpolarised
and therefore triplet supercurrents are able to carry a net spin. Spin-triplet pairs
have been detected in SC/FM hybrids as long-range correlations in a FM\cite{7,34,35}, through
critical temperature measurements of SC/FM multilayers\cite{36,37,38,39}, supercurrents in SC/FM/SC
Josephson junctions\cite{40,41,42,43,44,45} and density of state measurements\cite{46,47,48,49}. At a homogeneously
magnetized SC/FM interface, spin-zero triplet pairs form through a spin-mixing process. Spinzero
triplet pairs decay on the same length scale as singlet pairs which is typically a few
nanometres in a FM. In the presence of magnetic inhomogeneity\cite{6,7,8,34,35,50} and/or
appropriate spin-orbit coupling\cite{51,52,53,54} at a SC/FM interface, the spin-zero triplet pairs
convert to spin-polarized triplet pairs as illustrated in Fig.~\ref{fig2}(a). Spin-polarized triplet pairs
propagate through a FM over the triplet coherence length, which is tens nm and theoretically
limited to the spin-flip length of the FM.

Evidence for pure spin supercurrents in a SC were first shown by K.-R. Jeon et. al.\cite{55} through
spin-pumping experiments using Pt/Nb/Py/Nb/Pt multilayers in the superconducting state.
Here the effective Gilbert damping (which is proportional to the spin-current density) of the
precessing Py (Ni$_{80}$Fe$_{20}$) magnetization increased below the superconducting transition
temperature, indicating that the spin sink efficiency of Pt increased as a superconducting gap
in Nb opened as schematically illustrated in Fig.~\ref{fig2}(b). The explanation is that the strong spinorbit
coupling in Pt induces an evanescent spin-orbit interaction in Nb which, in conjunction
with an induced magnetic exchange in Pt from Py through Nb\cite{56,57}, creates triplet channel
through superconducting Nb. Since this equal-spin density of states channel in Nb offers a
pathway for transferring spin angular momentum, the spin current at the Py/Nb interface is
additionally transferred and absorbed in Pt, resulting in an enhanced spin current density
above the normal state. Although further work is required in order to establish the spin
transfer mechanism(s), this work provided compelling evidence for spin supercurrents.

As a result of the interplay between spin supercurrents and the absolute phase of the
superconducting state, spin supercurrents should be considered as more than just a low
dissipation equivalent of normal state spin currents. Nevertheless, the generation of nondissipative
spin currents without dissipative charge currents is of great importance for low
energy computing applications. Even under the unlikely assumption of 100\% spin polarisation,
the current densities achievable in the superconducting state have not yet reached the
necessary magnitudes required for conventional spin-transfer torque switching. Therefore,
the ability to exploit spin supercurrents is dependent on other interactions and phenomena,
which are only just being explored. Moreover, fundamental parameters such as the spinpolarisation,
the conversion rate from singlet-to-triplet pairs, decay lengths and times are
experimentally unclear and require intensive study both theoretically and experimentally.

\begin{figure}
\centering
\includegraphics[bb=0 0 725 359, scale=0.23]{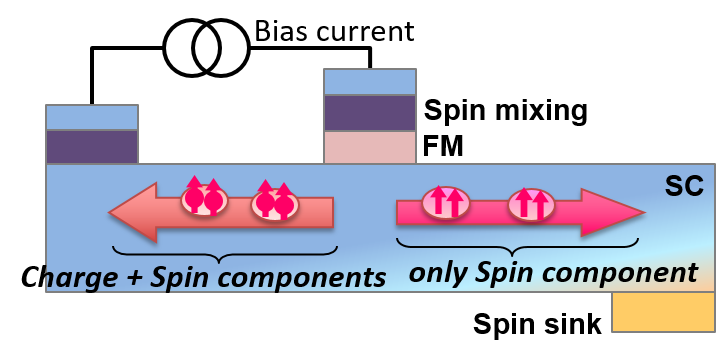}
\caption{Possible lateral circuit for pure spin supercurrents. In the right-hand half of the
superconducting channel which is out of the loop, the supercurrent only with the spin
component can be driven, namely a pure spin supercurrent.}
\label{fig3}
\end{figure}

The discovery of pure spin supercurrents in a conventional SC offers the potential for the
separation of spin and charge components in the superconducting state, which is the analogue
of non-local spin transport in normal state lateral devices. In the case of a diffusive current,
the current is driven by a gradient of electrochemical potentials. Since the spin supercurrent
is carried by coherence Cooper pairs we can expect that a gradient of absolute phase of the
superconducting condensate to act as an electrochemical potential. This alters the structures
required to generate this spin and charge separation of a triplet state. A structure to test this
is a SC/spin-mixer/FM/SC/spin-mixer/SC structure as shown in Fig.~\ref{fig3}. Here, the spin-mixing
layers perform the necessary singlet-to-triplet pair conversion process, and the central FM
spin-filters the spin-polarized triplet pairs. Although the charge component of the injected
spin supercurrent flows in accordance with the conventional coherence conditions, the 
equilibrium current-phase relations for spin and charge components will be different\cite{58,59}.
In other words, it may be possible to have a non-zero spin supercurrent while the phase
difference between the two superconductors is zero. Hence a spin accumulation without
charge flow (i.e. a pure spin supercurrent) could appear in the region away from the junction,
provided that a suitable spin-sink or spin-source is created. In contrast to the normal state,
the transmission range of this pure spin supercurrent is primarily limited not by the spin
diffusion length but the coherence length of the induced triplet state, which appears to be
comparable to the singlet coherence length. Even with this restriction, superconductors such
as Al are ideal for realising this type of lateral device due to its long coherence length of
hundreds of nanometres\cite{60,61}.

\begin{figure}
\centering
\includegraphics[bb=0 0 1384 343, scale=0.23]{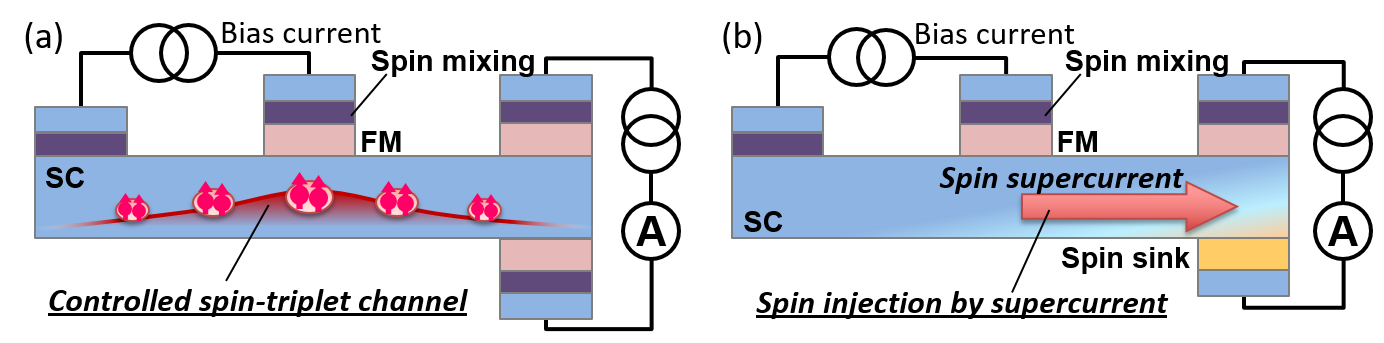}
\caption{Possible lateral circuit for nonlocal control via spin supercurrent. The spin supercurrent
induced by left-hand loop can control a supercurrent in the right-hand loop through (a)shared
superconducting channel or (b)strong spin-orbit material (spin sink).}
\label{fig4}
\end{figure}

We suggest two possible setups that involve spin supercurrents to create new functionality
for superconducting circuits. Firstly, if the structure in Fig.~\ref{fig3} is extended as shown in Fig.~\ref{fig4}(a)
to share the channel of spin supercurrent with another circuit, the magnitude of the charge
supercurrent in the right-hand loop may be controlled by the sign of the current in the left
loop. The spin supercurrent in the left loop determines the spin polarisation of the triplet state
in the SC channel. This naturally restricts the spin supercurrent in the right loop, depending
on the sign of the current and magnetisation of the electrodes. The magnitude of the critical
current in the right-hand loop may thus provide an experimental method to detect a pure spin
supercurrent which flows non-locally through the superconducting layer. A related scenario is
described theoretically in Ref.~\onlinecite{62} where a spin-polarized supercurrent is assumed to propagate
through an intrinsic triplet (p-wave) condensate.

An alternative coupling arrangement is shown in Fig.~\ref{fig4}(b). Here, a supercurrent flowing in the
left loop pumps spin into the spin sink which is the high spin-orbit material forming part of the
right-hand loop. The effective exchange field induced by spin accumulation in this material
acts with the spin-orbit coupling to create a triplet channel in the right-hand loops which only
has one conventional spin mixer. As a result, the spin supercurrent in the right-hand loop can
be controlled by that in the left-hand-loop. We would like to stress that, in the structure
shown in Fig.~\ref{fig4}(b), the injecting spins into the spin-orbit materials are not necessary to be
generated from supercurrent. The conventional spin injection or spin-polarised quasiparticles
are also available to induce an effective exchange field. There must be some
dissipation due to the spin flow in the spin-orbit materials, but this could be small.

There are a few key underlying assumptions in these proposals. In the spin supercurrent, the
charge and spin components must be manipulated separately -- i.e. the spin polarisation of the
spin supercurrent is determined without regard to the flow of the charge component in the
superconductor. In Fig.~\ref{fig4}(b), assuming that the triplet proximity channel can carry only Cooper
pairs, the conversion in spin current between condensates and quasi-particles at the contact
between SC and a high spin-orbit materials is inevitable to be absorbed or extracted.
Moreover, the injected spin accumulation in the spin-orbit materials has to act as an effective
exchange field. Experimental results\cite{55} show some possibilities, but there has so far been
no detailed theoretical exploration of them.

\begin{acknowledgments}
This work is supported by the EPSRC Programme Grant ``Superspin'' (EP/N017242/1) and the
JSPS Programme ``Fostering Globally Talented Researchers'' (JPMXS05R2900005).
\end{acknowledgments}


\end{document}